# Fingerprinting the magnetic behavior of antiferromagnetic nanostructures using remanent magnetization curves

M. J. Benitez,[1,2,*] O. Petracic,[1,†] H. Tüysüz,[2] F. Schüth,[2] and H. Zabel[1]

[1]Institut für Experimentalphysik/Festkörperphysik, Ruhr-Universität Bochum, D-44780 Bochum, Germany

[2]Max-Planck Institut für Kohlenforschung, D-45470 Mülheim an der Ruhr, Germany

Antiferromagnetic (AF) nanostructures from $Co_3O_4$, CoO and $Cr_2O_3$ were prepared by the nanocasting method and were characterized magnetometrically. The field and temperature dependent magnetization data suggests that the nanostructures consist of a core-shell structure. The core behaves as a regular antiferromagnet and the shell as a two-dimensional diluted antiferromagnet in a field (2d DAFF) as previously shown on $Co_3O_4$ nanowires [Benitez *et al*., Phys. Rev. Lett. **101**, 097206 (2008)]. Here we present a more general picture on three different material systems, i.e. $Co_3O_4$, CoO and $Cr_2O_3$. In particular we consider the thermoremanent (TRM) and the isothermoremanent (IRM) magnetization curves as "fingerprints" in order to identify the irreversible magnetization contribution originating from the shells. The TRM/IRM fingerprints are compared to those of superparamagnetic systems, superspin glasses and 3d DAFFs. We demonstrate that TRM/IRM vs. *H* plots are generally useful fingerprints to identify irreversible magnetization contributions encountered in particular in nanomagnets.

## I. INTRODUCTION

Magnetic nanostructures hold the potential for numerous applications, e.g., in magnetic data storage,[1,2] logic devices,[3-5] sensors[6] or bio-medical applications.[7,8] Usually a large variety of possible magnetic behaviors can be encountered depending on several factors like the material, the type of system (ferromagnet, ferrimagnet, antiferromagnet, etc.), interactions, sizes and shapes. This makes it often difficult to distinguish intrinsic physical properties of interest from mere artifacts. Sometimes complex superpositions of different behaviors occur hampering a unique interpretation. Also finite-size effects may create additional contributions or effects. E.g. an ideal antiferromagnet (AF) is expected to show zero magnetization in remanence. However, nanosized AF structures often show an excess magnetization due to the increased surface contribution. Therefore, the necessity for a characteristic magnetic 'fingerprint' arises so that different systems can be classified and distinguished.

In this article we aim to address two points: First, we generalize the previously observed behavior[9] by investigating and comparing three different AF materials, i.e. $Co_3O_4$, CoO and $Cr_2O_3$ in a network-like structure. Second, particular attention is drawn onto the TRM/IRM vs. *H* plots, which can serve as magnetic fingerprints to identify the irreversible magnetization contributions often encountered in nanosized systems. E.g. in AF nanostructures irreversible contributions are mainly due to the shell. This article is organized as follows. The experimental details are introduced in Sec. II. In Sec. III results and discussion of magnetization vs. temperature, magnetization vs. field and TRM/IRM plots are given. A summary and conclusions are presented in Sec. IV.

**A. Antiferromagnetic nanostructures**

Already Louis Néel discussed the effects of uncompensated surface spins in AF nanoparticles.[10] Many further studies picked up this question in order to clarify their underlying properties. Several studies suggest a spin-glass or cluster-glass-like behavior of the surface spins due to frustrations in the interactions.[11-14] Other studies propose thermal excitation of spin-precession modes,[15] or finite-size induced multi-sublattice ordering.[16] A number of publications describe the magnetic behavior in terms of an interaction between an AF core and a ferromagnetic-like shell.[12,13,17] Several studies explain the results in terms of weak ferromagnetism.[18,19] However, a precise understanding of the nature of the surface contribution has remained open. Recently we showed that for AF $Co_3O_4$ nanowires the magnetic behavior can be clearly described in terms of a core-shell system, where the core behaves as a regular AF and the shell as a 2d DAFF system.[9]

**B. Magnetic fingerprints**

Probably the most familiar 'fingerprint' of magnetic systems is the hysteresis loop $M(H)$. Hysteresis loops in ferromagnetic (FM) and ferrimagnetic systems are usually characterized by a non-linear $M(H)$ curve and irreversibilities upon field cycling (viz. 'open loop'). AF systems -in contrast- usually show a linear and closed hysteresis with often very large saturation fields (> 10 T). AF-FM composite systems may show the exchange bias (EB) effect [17, 20-22]. The EB, which results from the interaction between an AF with a FM via a common interface, manifests itself by a displacement of the hysteresis loop along the field axis after the system is cooled in a magnetic field below the Néel temperature of the AF. In practice often such a loop shift is taken as 'fingerprint' for any EB phenomena encountered in a sample. Or, it has also been shown that by studying the shape of the hysteresis loops on submicron circular nanomagnetic dots it is possible to identify the underlying spin structure.[23] i.e. a regular FM-like loop indicates a single-domain behavior, whereas a loop with a collapsed central part is characteristic for a vortex state.[23]

Also first-order reversal curve (FORC) diagrams are a useful tool to characterize magnetic systems with respect to their reversal behavior.[24,25] A FORC is measured after saturating the sample in a positive applied field. The applied field is lowered to the so-called reversal field $H_R$. Then, the FORC is the resulting magnetization curve when the field is increased until a field $H$. The magnetization at the applied field $H \geq H_R$ on a FORC with reversal field $H_R$ is denoted by $M(H, H_R)$. After computing the mixed second order derivative $\rho(H, H_R) = -(1/2)[\partial^2 M / \partial H \partial H_R]$ and changing variables to $H_c = (H - H_R)/2$ (local coercivity) and $H_b = (H + H_R)/2$ (local bias) one arrives at the "FORC distribution" $\rho(H_b, H_c)$, which is usually displayed as a 2D false color plot.[24,25] An example is the clear difference found between diagrams of a random-field Ising model (RFIM) and the Edwards-Anderson Ising spin glass (EASG). In the EASG case the FORC diagrams are characterized by a marked horizontal ridge, indicative of a broad range of effective coercivities in the system, but narrow range of biases. However, in a RFIM the FORC diagrams display a well-developed vertical feature reflecting a rather narrow range of effective coercivities and a broad range of biases.[24]

A fingerprinting method probing specifically the dynamic behavior is the so-called Cole-Cole plot.[26] The measurements are performed by applying a small oscillating magnetic field with driving frequency *f*, superimposed onto a constant magnetic field. The real and imaginary part of the ac susceptibility is the in-phase and out-of-phase component of the recorded time-dependent magnetization response. The ac susceptibility is measured as a function of the ac frequency, i.e. $\chi'(f)$ and $\chi''(f)$, at a constant temperature and magnetic field. The Cole-Cole plot is then obtained by plotting the imaginary part $\chi''$ against the real part, $\chi'$ and thus eliminating the *f*-dependence. One arrives at various shapes of $\chi''(\chi')$-curves depending on the specific system. The most simple feature is a semicircle ('Debye-semicircle') signifying the presence of just *one* relaxation time in the system. It has been demonstrated that e.g. superparamagnetic systems can be distinguished from superspin glass or superferromagnetic by their Cole-Cole plots.[27]

Another fingerprinting method employs the measurement of the remanence (the remaining magnetization after the applied magnetic field is reduced to zero). This is particularly important in systems suitable for magnetic recording purposes, where magnetic interactions can have a strong influence on the signal-to-noise ratio.[28] Applying a dc magnetic field, it is possible to measure three relevant remanent magnetization curves namely the thermoremanent (TRM), the isothermoremanent (IRM) and the dc demagnetization (DCD) curve. To measure the TRM, the system is cooled in the specified field from a high temperature down to the measuring temperature, the field is then removed and subsequently the magnetization is immediately recorded, whereas to measure the IRM the sample is cooled in zero field from high temperature down to the measuring temperature, the field is then momentarily applied, removed again and then the remanent magnetization is immediately recorded. The DCD is measured after the sample is cooled in zero field from high temperature down to the measuring temperature, where the sample is first saturated in one field direction. The field is then momentarily applied in the opposite direction, removed again and then the remanent magnetization is recorded. One example of the use of remanence curves is the well known $\Delta M$ method where $\Delta M(H)$ curves are obtained from DCD and IRM procedures. The $\Delta M$ is defined by $\Delta M(H) = M_{DCD}(H)/M_R - [1 - 2\,M_{IRM}(H)/M_R]$, where $M_R$ is the saturation remanence, and is often used to characterize magnetic interactions between nanostructures.[29] If the interparticle coupling is dominated by exchange interaction, $\Delta M$ is positive, whereas for interactions of dipolar type, $\Delta M$ becomes negative.[29]

In this article we draw the attention to another fingerprinting method based on TRM/IRM vs. field $H$ measurements. This method has already been employed previously in the context of random magnets e.g. DAFF systems,[30] but is yet unknown as a tool for nanomagnetic systems. TRM/IRM plots represent a useful method to identify the nature of the *irreversible* magnetization contributions. Reversible contributions become zero in the TRM/IRM plot. E.g. an ideal AF bulk system is expected to show both zero TRM and zero IRM for all fields and temperatures. Here we employ the TRM/IRM vs. $H$ plots to separate or 'enhance' the contribution of the shells of AF nanostructures. These TRM/IRM fingerprints can then be compared to other systems, e.g. superparamagnets, spin glasses and 3d DAFF systems.

## II. STUDIED SYSTEMS AND EXPERIMENTAL DETAILS

We have studied three different AF systems, $Co_3O_4$, CoO and $Cr_2O_3$. Bulk $Co_3O_4$ has a 'direct spinel' structure where the $Co^{3+}$ and $Co^{2+}$ ions are in the octahedral and the tetrahedral sites, respectively.[31,32] In bulk $Co_3O_4$ the magnetic transition from paramagnetic state to AF occurs at 40 K. The second system, CoO, has a sodium chloride structure in the paramagnetic state. Below the Néel temperature, $T_N$ = 290K CoO becomes tetragonally distorted with c/a<1.[33] The third material, $Cr_2O_3$ is chosen because of its characteristic spin-flop phase.[34] $Cr_2O_3$, which is a uniaxial antiferromagnet, crystallizes in a corundum structure ($R\bar{3}c$). Below the Néel temperature ($T_N$ = 307 K),[35] in zero magnetic field, the $Cr^{3+}$ spins align antiferromagnetically along the [111] easy axis, whereas at the spin-flop transition the spins are reoriented in the basal plane maintaining the AF order.[36] Spin-flop field values for bulk $Cr_2O_3$ correspond to 60 kOe at 4.2 K.[34] With decreasing particle size the spin-flop field $H_{SF}$ decreases. Values of $H_{SF}$ = 10 kOe at 5 K were measured for nanoparticles with ellipsoidal shape with the major axis of approximately 170 nm and the minor axis 30 nm.[36] A further reason for choosing this system is these field values that are in the usual experimentally accessible range and it is thus possible to study TRM/IRM curves close and below the spin-flop transition for $Cr_2O_3$ nanostructures.

All nanostructures were prepared via the so-called nanocasting method (Fig. 1(d)).[37,38,39] Detailed description of the synthesis and structural characterization of these AF materials has been reported previously.[39,40] In particular, the resulting AF materials were characterized in detail at different synthesis steps during the templating route by transmission and scanning electron microscopy and by powder X-ray diffraction. Electron microscopy investigations show well ordered nanostructures, whereas X-ray diffraction patterns confirm a single $Co_3O_4$, CoO or $Cr_2O_3$ phase.

High resolution scanning electron microscopy (HRSEM) images of the samples were taken using a Hitachi S-5500 ultra-high resolution cold field emission scanning electron microscope operated at 30 kV. All samples were prepared on lacey carbon films supported by a copper grid. The obtained images were analyzed using the Scandium 5.0 software package from Soft Imaging System GmbH. Figure 1 shows the HRSEM images of (a) $Co_3O_4$, (b) CoO and (c) $Cr_2O_3$ cubic ordered AF nanostructures with 8 nm diameter of the oxide struts forming the network. Magnetometry measurements of the samples were performed using a Quantum Design MPMS5 superconducting quantum interference device (SQUID) magnetometer in applied magnetic fields up to 50 kOe.

## III. RESULTS AND DISCUSSION

### A. Magnetization vs. temperature curves

Fig. 2 shows $M$ vs. $T$ curves after zero field cooling (ZFC) and after field cooling (FC) measured on cubic ordered $Co_3O_4$ nanostructures at two applied fields, 40 kOe (a) and 50 Oe (b). In each case the sample was cooled from room temperature down to 5 K. For a *regular* bulk AF a peak both in the ZFC and FC curve is expected, when the field is applied along the anisotropy direction. The inflection point left to the peak position marks the critical temperature $T_c(H)$, with $T_c(0) = T_N$.[41] Instead, often in literature the peak position itself is used to mark the critical temperature $T_c(H)$.[12,13,42-44] Here we adopt the inflection point definition. For a small field of 50 Oe the inflection point corresponds to $T_c(50\ \text{Oe}) \approx T_N = 27$ K. It should be noted that the Néel temperature is reduced compared to the bulk value of $T_N = 40$ K due to the finite size effect[42,45] and not to dilution effects in the core.

Next, we measured $M$ vs. $T$ curves at 40 kOe. One finds basically no change in the inflection point compared to the curve measured at 50 Oe. This matches with the previous findings on $Co_3O_4$ nanowires.[9] In most AF systems the field dependence of the critical phase boundary is very small in the range of the usually accessible experimental field values, i.e. $H < 50$ kOe. Therefore, we can conclude that the cubic ordered $Co_3O_4$ nanostructures consist of *AF* *ordered cores*, which behave purely AF. Note that the Néel temperature confirms the single phase structure as also obtained from the X-ray diffraction studies.[39]

The $M$ vs. $T$ curves measured both at 50 Oe and 40 kOe for $Co_3O_4$ nanostructures show a splitting (bifurcation) of the FC and ZFC magnetization below a temperature $T_{bf}$. These results are in agreement with previous studies on $Co_3O_4$ nanowires.[9] The splitting is due to an irreversible magnetization contribution and has been attributed to the presence of a 2d-DAFF shell of the nanowires.[9] The irreversible contribution can be better seen by plotting the difference, $\Delta M = M_{FC} - M_{ZFC}$ (Fig. 2, insets). The $\Delta M$ curves reach zero at $T_{bf} = 25$ K (for FC in 40 kOe) and $T_{bf} = 27$ K (for FC in 50 Oe).

These findings can be extended to other AF systems. Fig. 3 shows $M$ vs. $T$ curves measured for CoO (a,b) and $Cr_2O_3$ (c,d) nanostructures after ZFC and after FC, measured at two applied fields, i.e. 40 kOe (a,c) and 50 Oe (b,d). In each case the sample was cooled down from 400 K to 5 K. Qualitatively a similar behavior is found as in the case of $Co_3O_4$ nanostructures, i.e. a peak in ZFC curve with the inflection point marking the Néel temperature $T_N$ and a splitting of ZFC-FC curves below $T_{bf}$. We find that for both, CoO and $Cr_2O_3$ nanostructures again no field dependence exists of the inflection point in the ZFC curve. In the case of CoO this is $T_N = 260$ K and in the case of $Cr_2O_3$ $T_N = 300$ K. From this finding we conclude that the CoO and $Cr_2O_3$ nanostructures consist of *AF ordered cores*, which behave purely AF. The reduced Néel temperatures are again attributed to finite size effects.

### B. Magnetization vs. field hysteresis curves

Magnetization hysteresis loops at 5 K after ZFC and FC on cubic $Co_3O_4$ nanostructures are shown in Fig. 4 (a). One observes a small coercivity of 78 Oe in the ZFC curve and a virtually linear shape in the field range used, $|H| < 40$ kOe. This matches well with the previous results found on $Co_3O_4$ nanowires.[9] The overall linear behavior is due to the regular AF nanowire cores, while the irreversible contribution (viz. the loop opening) has been attributed to the 2d-DAFF shells.[9] The hysteresis curve measured after FC in 40 kOe displays an enhancement of the coercive field to 146 Oe and a vertical shift to larger $M(H)$ values. This also matches with previous results on $Co_3O_4$ nanowires[9] and with other hysteresis loops on DAFF systems.[30]

Fig. 4 (b) shows hysteresis loops at 5 K after ZFC and FC on cubic CoO nanostructures. The $M$ vs. $H$ curve after ZFC is completely closed (viz. does not show any hysteretic behavior). The corresponding curve after FC in 40 kOe displays an enhancement of the coercive field to 264 Oe and a vertical shift to larger $M(H)$ values similar to the $Co_3O_4$ nanostructures.

Results for the cubic $Cr_2O_3$ nanostructures are depicted in Fig. 4 (c). The deviation from the linearity of the ZFC $M$ vs. $H$ is attributed to a spin-flop transition.[36] The corresponding $M$ vs. $H$ curve after FC shows a similar deviation from the linearity accompanied by a shift in the hysteresis loop as in the cases discussed before.

Magnetization hysteresis loops for $Cr_2O_3$ nanostructures after ZFC at different temperatures 20 K, 70 K and 200 K are shown in Fig. 5(a). One notices that at 20 K and 70 K there is still a deviation from linearity in $M(H)$, whereas at 200 K the magnetization shows a linear dependence of $H$ as expected for AF systems. ZFC and FC magnetization hysteresis loops at 200 K are shown in Figure 5(b). A small coercivity in the ZFC curve and a shifted hysteresis after FC in 40 kOe is obtained.

### C. Magnetization curves at remanence

In this section we discuss the TRM/IRM magnetization curves as a function of field and temperature. It is important to note that the TRM and the IRM magnetization curves probe two different magnetic states of the system. The TRM probes the remanent magnetization in zero field after freezing-in a certain magnetization in an applied field during FC. However, the IRM probes the remanent magnetization in zero field after ZFC (in a demagnetized state) and then magnetizing the system at low temperatures, probing only those spins which are still switchable. Thus, it is expected that systems with a non-trivial *H-T*-phase diagram exhibit characteristically different TRM and IRM curves. Fig. 6 shows the TRM and IRM curves as function of magnetic field (a) of the canonical spin glass (SG) system AuFe adapted from Ref. 46, (b) of superparamagnetic (SPM) Fe particles, with a mean diameter of 3nm embedded in a alumina matrix adapted from Ref. 47, (c) of bulk-DAFF system, $Fe_{1-x}Zn_xF_2$ adapted from Ref. 30, and (d) of $Co_3O_4$ nanowires.[9]

It has long been known that the magnetic behavior of a SG system strongly depends on whether it is cooled in a field or not.[48] Therefore, characteristic differences between TRM and IRM are observed. Theoretical studies using Monte-Carlo simulations show that the remanent magnetization curves depend on the final temperature and the field which was applied initially. Higher values of TRM in comparison with IRM are expected due to the fact that TRM starts from a high magnetization. TRM grows linearly with the field and exhibits a characteristic peak for field energies of the order of the interaction energy ($\approx k_B T_f$).[49] The interaction field is assumed to be negative and increases as the field increases.[50] The IRM increases relatively strongly with increasing field and meets the TRM curve at moderate field values, where both then jointly saturate. This scenario is observed in the AuFe SG system [Fig. 6 (a)]. The TRM as a function of temperature decays linearly with temperature, whereas the IRM as a function of temperature has a maximum that is explained by the variation of the single cluster relaxation time with temperature.[49] Experimental studies from several other SG systems found in literature are in agreement with this theoretical approach.[49,51,52]

In a SPM system the remanence is related to the distribution of energy barriers in the system.[28] At a given measurement temperature and after removing the applied field, only the particles which are in the blocked regime will contribute to the remanent magnetization.[28] Theoretical[47] and experimental[47,53] studies on Fe particles in an alumina matrix show that in a system of non-interacting nanoparticles TRM increases with field and reaches saturation more rapidly than the IRM. The latter one increases relatively strongly with increasing field and meets the TRM curve where both then saturate [Fig. 6 (b)]. In contrast, 3d DAFFs are characterized by two interesting scenarios. Upon ZFC the system develops long range order, however upon FC the system breaks up into a metastable domain-state.[54] This behavior yields zero IRM for all fields and TRM which increases proportionally to $R^{-1}$, where $R$ is the domain size.

Next we show that the irreversible magnetization contribution can be independently probed by employing TRM and IRM vs. field. To measure the TRM, the system was cooled in the specified field from room temperature in the case of $Co_3O_4$ and 400 K in the case of CoO and $Cr_2O_3$ down to 5 K. Then the field was removed and the magnetization was recorded immediately. To measure the IRM, the sample was cooled in zero field from room temperature in the case of $Co_3O_4$ and 400 K in the case of CoO and $Cr_2O_3$ down to 5 K, the field was then momentarily applied (60 s), removed again and the remanent magnetization was recorded. Figure 7 shows the TRM/IRM vs. $H$ at 5 K for $Co_3O_4$ and CoO cubic ordered AF nanostructures. For $Co_3O_4$ we observe that the IRM stays at very small values even for fields up to 50 kOe, whereas the TRM curve shows a monotonic increase with a rounded maximum at $H \approx 40$ kOe. A maximum in the TRM is considered to be characteristic for a SG phase as discussed above. However, the hysteresis curves [Fig. 4(a)] do not support a SG scenario, because they would show a pronounced S-shape with significant loop opening.[48] Moreover, the small IRM signal and the shape of the curve as seen in Fig. 7 contradict both a SG and a SPM behavior.

3d DAFF systems are characterized by a zero IRM for all fields and a TRM which increases proportionally with the field.[30] The solid line in Figure 6 (c) is a fit to the TRM data according to the power law, TRM $\propto H^{v_H}$, with $v_H = 3.05$.[26] The TRM of the $Co_3O_4$ nanowires displays also a monotonically increasing curve, however with $v_H < 1$. The dimensionality and the finite size of the DAFF system play a crucial role in the TRM/IRM behavior and in particular the field dependence of the TRM so that a 2-dimensional finite-size DAFF system is likely to show a TRM vs. $H$ behavior as found in the $Co_3O_4$ nanowires. Temperature dependent magnetization studies confirm the dimensionality of the shell as a 2d DAFF.[55]

For CoO with cubic structure at $T = 5$ K one observes that the TRM has qualitatively similar behavior to that found for $Co_3O_4$, however the IRM is zero even for large fields up to 50 kOe. This hints to a more pronounced DAFF type behavior with less surface disorder. Figure 8 shows TRM/IRM vs. $H$ at 5 K and 200 K for $Cr_2O_3$ cubic ordered nanostructures. At 200 K we observe that the IRM stays at very small values even for fields up to 50 kOe, whereas the TRM curve shows a monotonic increase. This result is qualitatively similar to the TRM/IRM shown by $Co_3O_4$ and CoO, Note that the hysteresis loops at 200K support this scenario. At 5 K one finds that the TRM increases and reaches a maximum at 20 kOe. The IRM vs. $H$ increases and reaches a maximum at 35 kOe. This new feature could be related with the spin-flop phase being known to occur in $Cr_2O_3$. The reduced maximum of 20 kOe in the TRM compared with the 35 kOe in the IRM is likely a manifestation of the AF core together with a 2d DAFF shell.

Figure 9 shows the TRM (measured upon warming in zero field after FC in 40 kOe) vs. $T$ of (a) $Co_3O_4$, (b) CoO and (c) $Cr_2O_3$ nanostructures. One observes a characteristic temperature at which the TRM vanishes. It matches with $T_N$, which marks the ordering temperature of the AF cores, i.e. 27 K, 260 K and 300 K, respectively. The decay of TRM with increasing the temperature can be attributed to the frozen behavior of the 2d DAFF shell, which finally completely vanishes at $T_N$.

**IV. CONCLUSIONS**

In summary, our studies demonstrate the potential of TRM/IRM measurements to serve as a fingerprint to characterize the magnetic behavior of nanosystems. We have investigated three different AF systems, i.e. $Co_3O_4$, CoO and $Cr_2O_3$ nanostructures, which have been prepared by the nanocasting method from silica templates. Using SQUID magnetometry we have studied the magnetic behavior. Based on results from TRM/IRM vs. field of the AF systems discussed here, we can make the general observation that an increasing TRM and a small IRM signal are expected for AF nanostructures. Using TRM/IRM plots vs. field we can also confirm unambiguously a core-shell behavior consisting for all three systems of a regular AF core and a shell that magnetically behaves as two-dimensional diluted antiferromagnet in a field (2d DAFF) system.

**ACKNOWLEDGEMENTS**

We thank H. Bongard for the HRSEM images. The authors thank F. Radu and K. Westerholt for helpful discussions. One of the authors (M.J.B.R.) acknowledges support from the International Max-Planck Research School "SurMat".

**FIGURE CAPTIONS**

**FIG. 1.** HRSEM images of $Co_3O_4$ (a), CoO (b) and $Cr_2O_3$ (c) nanostructures with 8 nm crystallite size. The inset shows a schematic representation of the AF-DAFF core-shell structure. (d) Schematic of nanocasting method taken from Ref. [37] for the example of a hexagonal mesostructure.

**FIG. 2.** $M$ vs. $T$ curves after zero field cooling (ZFC) and after field cooling (FC) measured at two applied fields, i.e. 40 kOe (a) and 50 Oe (b) for $Co_3O_4$. The insets show $\Delta M = M_{FC} - M_{ZFC}$. The bifurcation temperature $T_{bf}$ is marked by an arrow.

**FIG. 3.** $M$ vs. $T$ curves after zero field cooling (ZFC) and after field cooling (FC) measured at two applied fields, i.e. 40 kOe (a,c) and 50 Oe (b,d) for CoO (a,b) and $Cr_2O_3$ (c,d) nanostructures, respectively. The insets show an enlarged view of $T_N$.

**FIG. 4.** $M$ vs. $H$ hysteresis curves at 5 K after ZFC and after FC of $Co_3O_4$ (a), CoO (b) and $Cr_2O_3$ (c) nanostructures, respectively. The insets show an enlarged view of the central part.

**FIG. 5.** (color online) $M$ vs. $H$ hysteresis curves of $Cr_2O_3$ nanostructures at 20 K (red open circles), 70 K (blue open triangles) and 200 K (solid black line) after ZFC(a) and $M$ vs. $H$ hysteresis curves at 200 K after ZFC and after FC (b). The inset shows an enlarged view of the central part.

**FIG. 6.** (color online). TRM and IRM vs. $H$ of the SG system AuFe(0.5%) adapted from Ref. 46 (a), of SPM Fe particles, with a mean diameter of 3 nm embedded in alumina adapted from Ref. 47 (b), of the DAFF system $Fe_{0.48}Zn_{0.52}Fe$, adapted from Ref. 30 (c) and of $Co_3O_4$ nanowires (NWs) at 5 K adapted from Ref. 9 (d).

**FIG. 7.** (color online). TRM (square black solid symbols) and IRM (square black open symbols) vs. $H$ for $Co_3O_4$ nanostructures at 5 K with a crystallite size of 8nm. TRM (circle red solid symbols) and IRM (circle red open symbols) vs. $H$ for CoO nanostructures at 5 K with a crystallite size of 8nm.

**FIG. 8.** (color online). TRM (square black solid symbols) and IRM (square black open symbols) vs. $H$ for $Cr_2O_3$ nanostructures at 5 K with a crystallite size of 8nm. TRM (circle red solid symbols) and IRM (circle red open symbols) vs. $H$ for cubic ordered $Cr_2O_3$ nanostructures at 200 K with a crystallite size of 8 nm.

**FIG. 9.** TRM vs. $T$ measured upon warming in zero field after FC in 40 kOe of $Co_3O_4$ (a) and CoO (b) nanostructures, respectively. TRM vs. $T$ measured upon warming in zero field after FC in 40 kOe from 400 K down to 5 K (black triangles) and from 400 K down to 200 K (red stars) of $Cr_2O_3$ nanostructures (c). The Néel temperature $T_N$ is marked by an arrow.

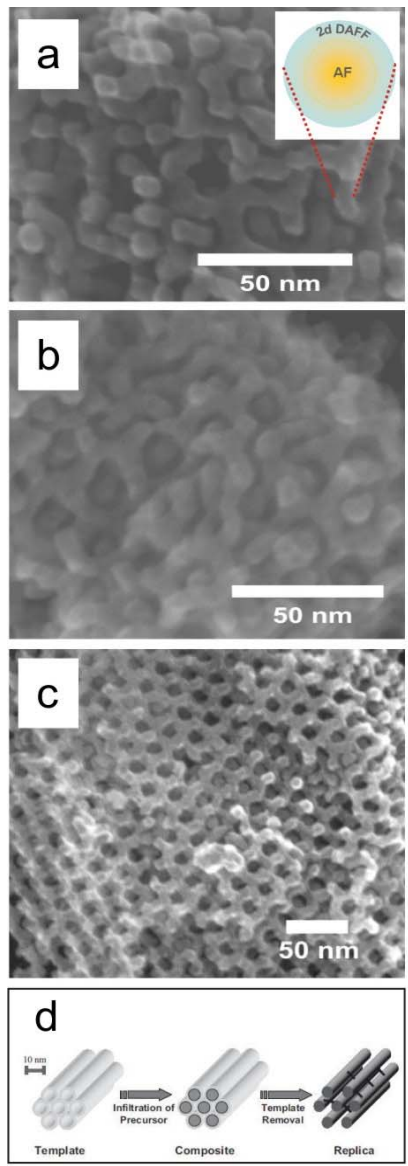

**Figure 1**

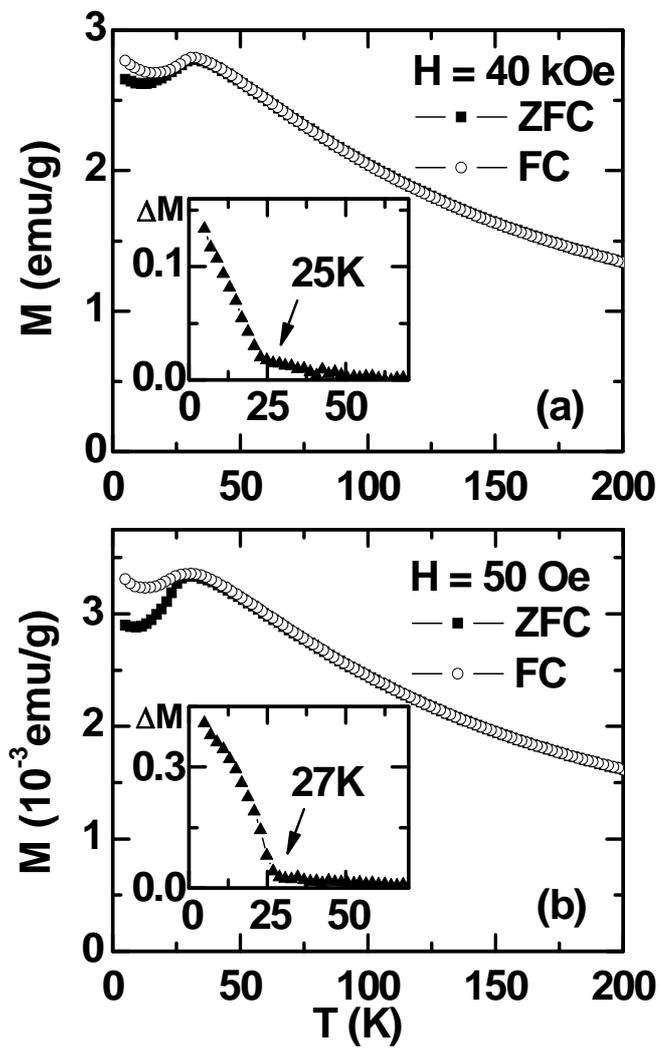

**Figure 2**

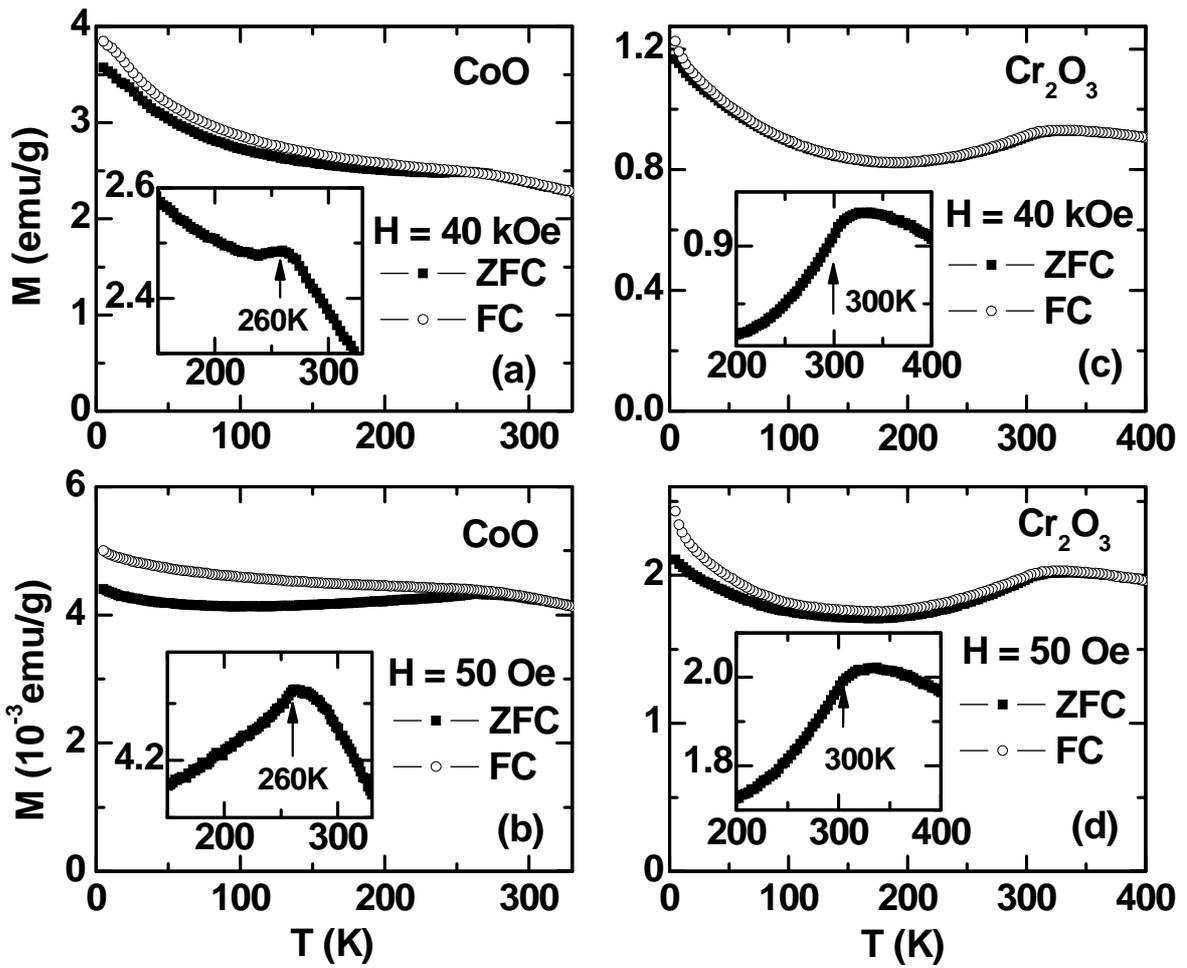

**Figure 3**

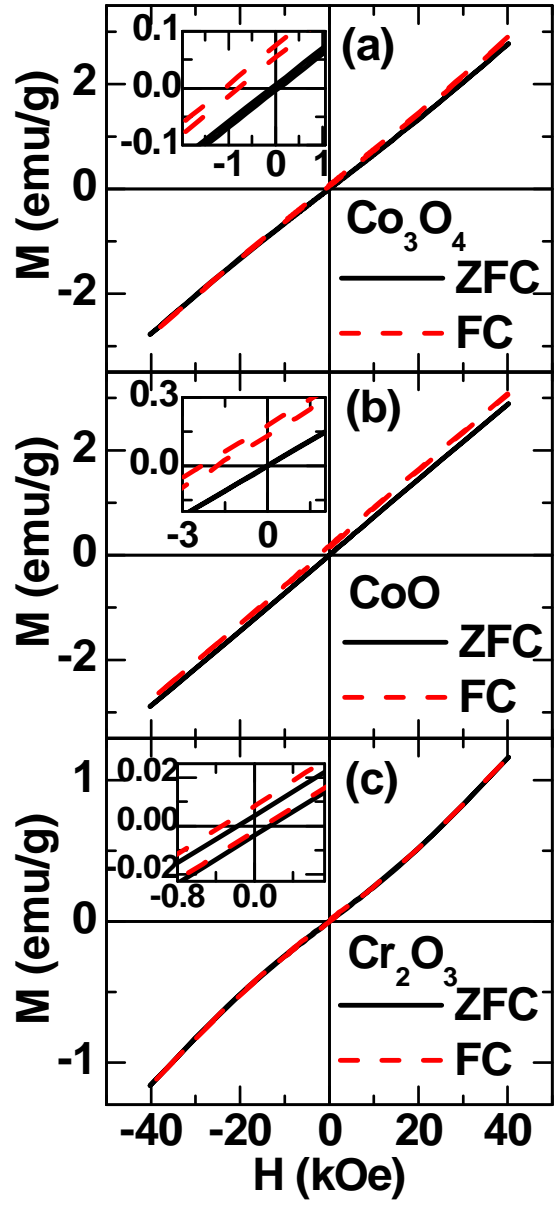

**Figure 4**

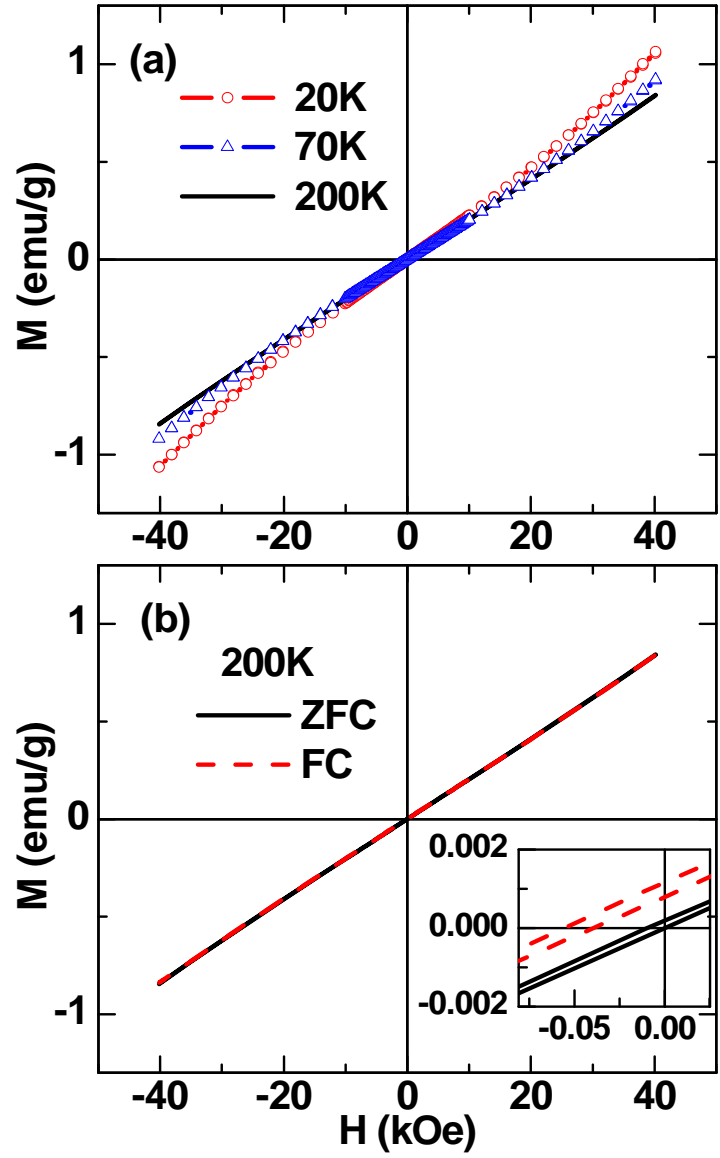

**Figure 5**

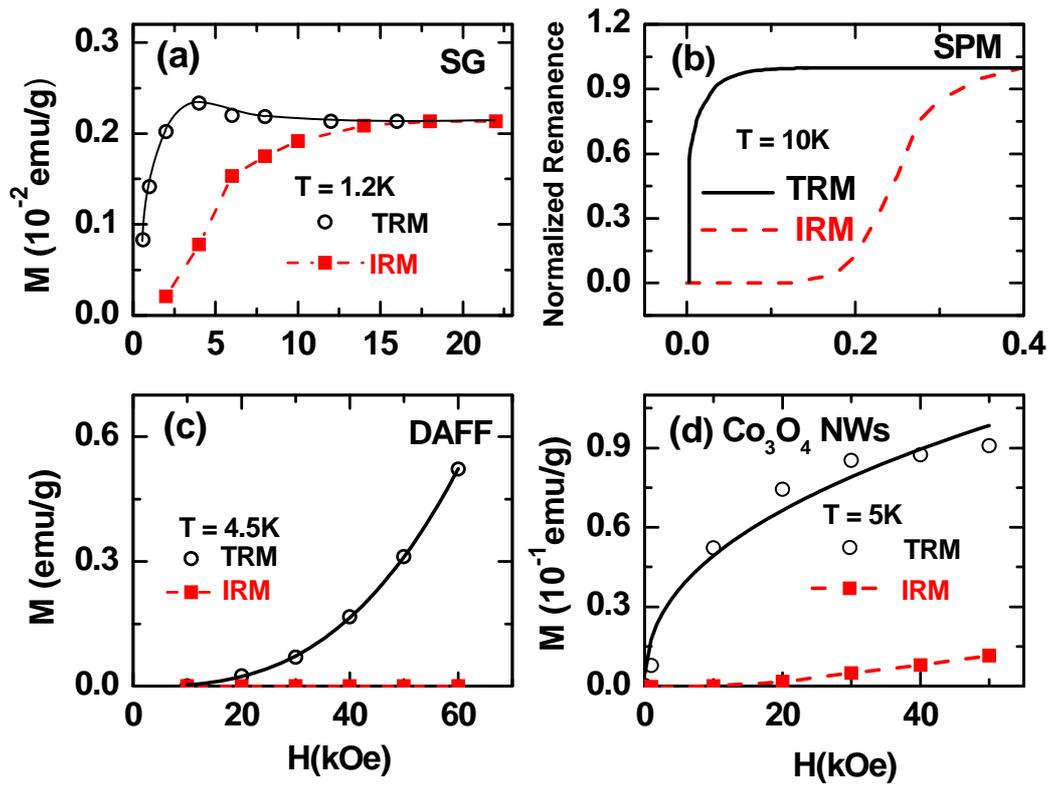

**Figure 6**

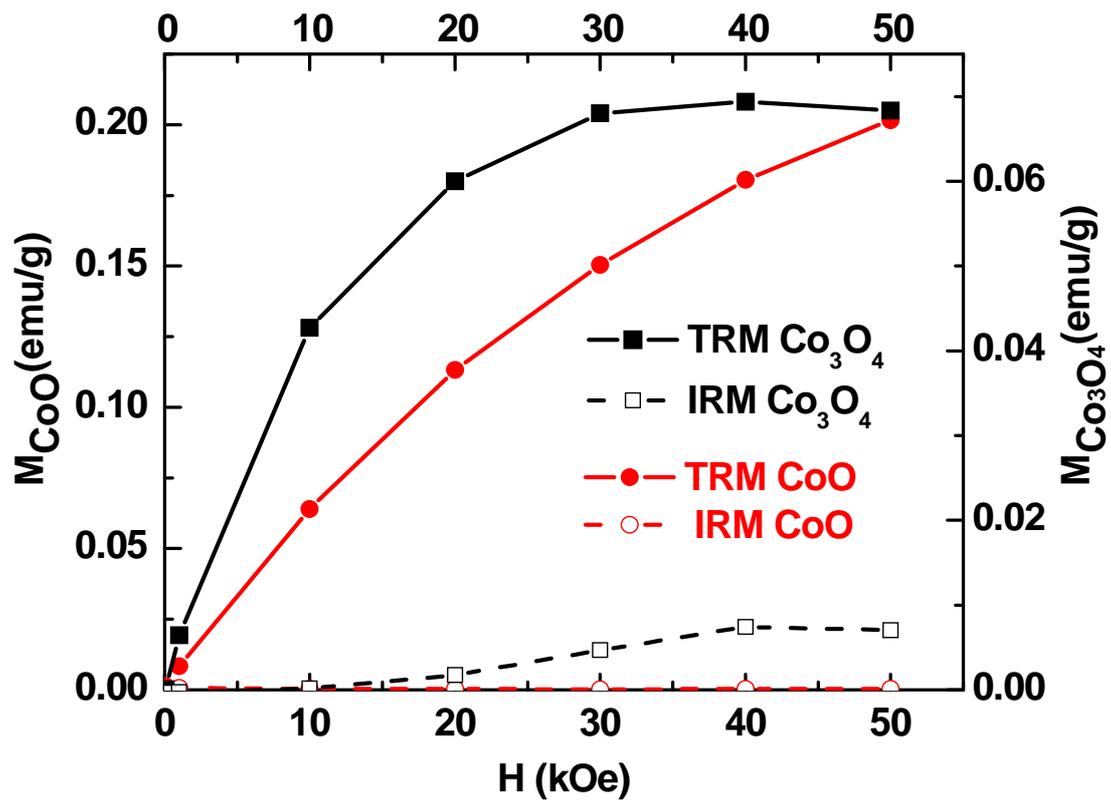

**Figure 7**

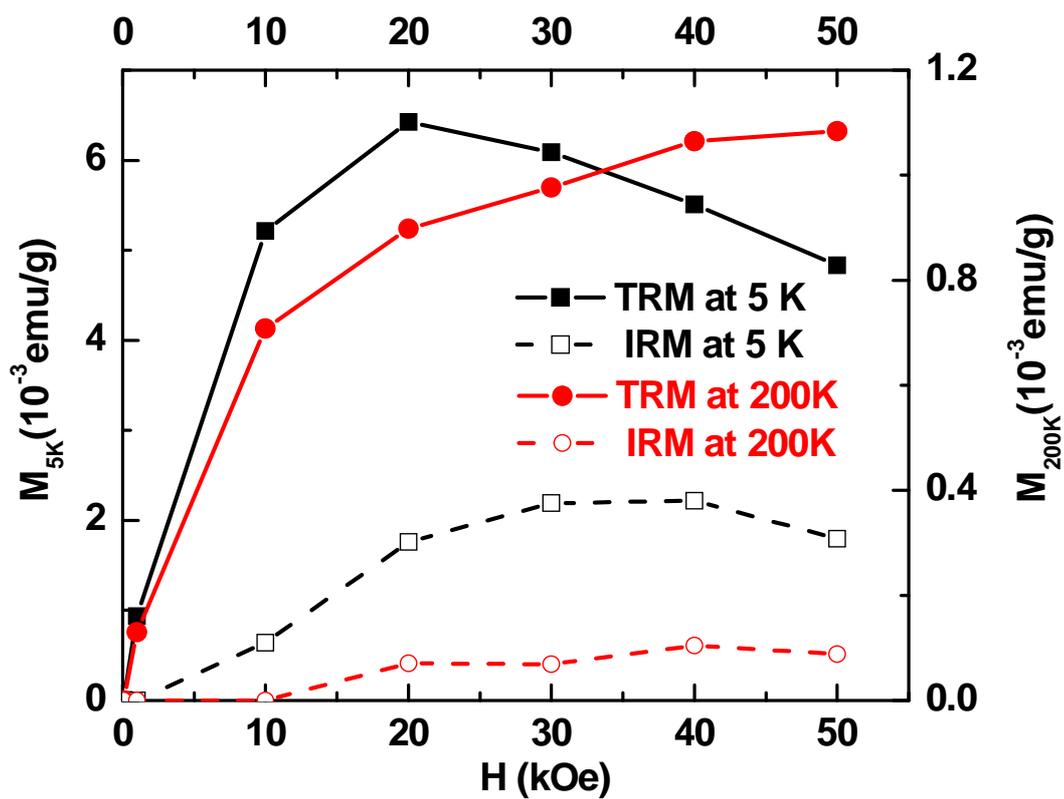

**Figure 8**

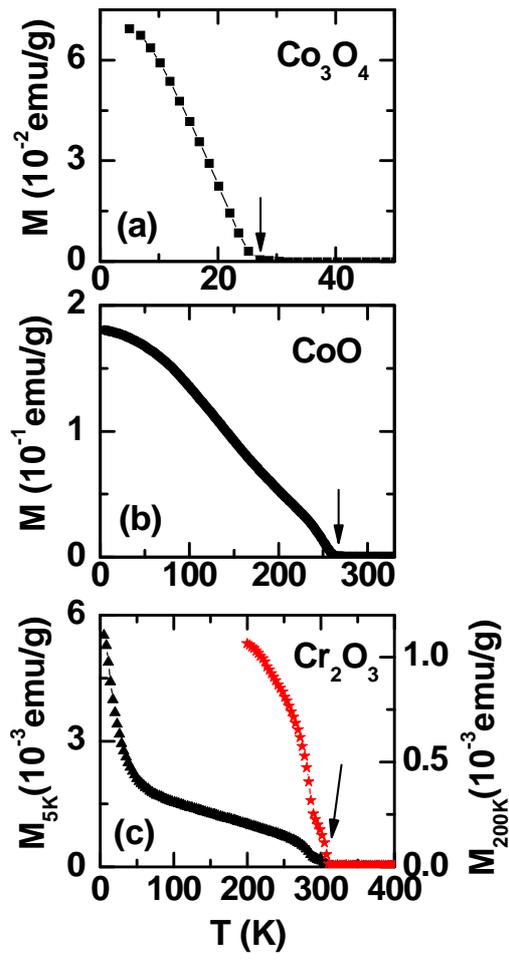

**Figure 9**